\documentclass[11pt,twoside]{article}
\usepackage{asp2006}
\usepackage{epsf}
\usepackage{psfig}
\usepackage{lscape}
\usepackage{txfonts}
\begin{document}

\title{The GMRT Radio Halo Survey and low frequency follow-up}

\author{T.~Venturi,$^1$ S.~Giacintucci,$^2$ R.~Cassano,$^1$ G.~Brunetti,$^1$
        D.~Dallacasa,$^3$ G.~Macario,$^1$ G.~Setti,$^3$ S.~Bardelli,$^4$
        and R.~Athreya$^5$}

\affil{$^1$INAF, Istituto di Radioastronomia, Bologna, Italy\\
       $^2$Harvard-Smithsonian Center for Astrophysics, Harvard, USA\\
       $^3$Dipartimento di Astronomia, Bologna University, Italy\\
       $^4$INAF, Osservatorio Astronomico di Bologna, Italy\\
       $^5$NCRA, TIFR, Pune University Campus, India}

\begin{abstract} 
The GMRT Radio Halo Survey, carried out at 610~MHz to investigate the
statistical properties of cluster radio halos in a complete cluster sample
selected in the redshift interval $z=0.2{-}0.4$, has significantly improved our
understanding of the origin of cluster radio halos and relics. Here we briefly
summarize the most relevant results of our investigation. A low frequency
follow-up is in progress with the  GMRT at 325~MHz and 240~MHz on the diffuse
sources and candidated found at 610~MHz. We briefly report some preliminary
results on these low frequency observations. Cluster radio halos with different
radio spectral properties have been unexpectedly found.
\end{abstract}

\section{Scientific background of the GMRT Radio Halo Survey}

Thermal and non-thermal components co-exist in clusters of galaxies. Beyond the
galaxies and the hot intracluster medium, diffuse extended (up and above the
Mpc) emission in the form of radio halos and/or relics is known nowadays in a
total of $\sim$ 40 clusters, which prove the existence of relativistic
particles ($\gamma \gg 1000$) and large scale magnetic fields (with intensity of
the order of the $\muup$G). The steep spectrum of these diffuse sources, with
$\alpha \sim 1.2 - 1.4$ (S$\propto \nu^{-\alpha}$) is a signature of their
synchrotron origin. Radio halos and relics do not have an obvious optical
counterpart, and they are not even the result of blending of individual
sources, but they rather permeate the cluster volume, similar to the thermal
hot gas responsible for the X-ray emission. These two classes of sources share
similar observational properties, such as very low surface brightness (of the
order of the $\muup$Jy arcsec$^{-2}$) and  steep synchrotron spectrum, while they
differ in the location within the cluster -- radio halos being centrally
located, in good coincidence with the ICM, relics being found in peripheral
cluster regions -- and in their polarization properties, i.e.\ radio halos are
unpolarized while relics show a high degree of fractional polarization (see
Feretti 2005 for a review). The longstanding crucial question concerning halos
and relics concerns their origin: the radiative life-time of the radiating
electrons is much shorter than the diffusion time necessary to cover the
cluster scale volumes, therefore their existence requires some form of particle
re-acceleration (see Cassano, present volume).

Among the variety of models proposed so far for the formation of giant (Mpc
scale) radio halos, the so-called ``re-acceleration model'', whereby electrons
are re-accelerated in-situ by MHD turbulence injected in the cluster volume by
merger events (Brunetti et al.\ 2001; Petrosian 2001), has received particular
attention. Starting from the re-acceleration model, a number of statistical
predictions (based on Montecarlo simulations) were made on the fraction of
clusters expected to develop a radio halo as a function of redshift and mass
(Cassano \& Brunetti 2005; Cassano, Brunetti \& Setti 2006, CBS06; Cassano,
this volume). Following those predictions, the bulk of giant radio halos is
expected in the redshift interval $z=0.2{-}0.4$, where $\sim 35$\% of massive
(i.e.\ $M > 10^{15}$ M$_\odot$) may host one.

The GMRT Radio Halo Survey was designed to test the statistical predictions
derived from the re-acceleration model, in particular the fraction of massive
clusters hosting a radio halo and its possible connection with the cluster
mass, and at a more general level to investigate the link between cluster
mergers and the presence of diffuse cluster sources, or lack thereof.

\begin{table}
\caption{Clusters with the GMRT at low frequency}
\label{ss_src}
\begin{tabular}{l c c c c l }
\hline
Cluster name & $z$ & Source type & $S_{\rm 610~MHz}$ & $\nu$ & Notes \\
             &     &             &      mJy          & ~MHz  &       \\
(1) & (2) & (3) & (4) &  (5) & (6) \\
\hline
A\,209   & 0.2060 & Giant Halo      &  24.0$\pm$3.6 & 325 & (a)     \\
A\,521   & 0.2475 & Relic+Giant halo&  41.9$\pm$2.1 & 325, 240 & (b)   \\
A\,697   & 0.2820 & Giant Halo      &  13.0$\pm$2.0 & 325 & (c)     \\
A\,781   & 0.2984 & Candidate relic &  32.0$\pm$2.0 & 325 & (d)     \\
A\,1682  & 0.2260 & Candidate halo  &   $\sim$ 44   & 240 & (d)     \\
A\,1300$^{\star}$  & 0.3075 & Giant Halo+Relic&      & 325 & (d) \\
A\,2744$^{\star}$  & 0.3066 & Giant Halo+Relic&      & 325 & (d) \\
RXCJ1314.4$-$2515  & 0.2439  & Halo + 2 Relics &     & 325, 240 & (a) \\
RXCJ2003.5$-$2323  & 0.3171 & Giant Halo      & 96.9$\pm$5.0& 325, 240& (c) \\
Z\,2661  & 0.3825 & Candidate Halo  &   $\sim$ 5.9  & 325 & (a)     \\
\hline \hline
\end{tabular}
Notes: $^{\star}$~halos known from the literature; (a)~data reduction still in
progress; (b)~Giacintucci et al.\ 2008 and Brunetti et al.\ 2008; (c)~imaging
and analysis completed; (d)~analysis in progress.
\end{table}

\section{Sample selection and summary of the results}

The galaxy cluster sample for our study consists of 50 clusters, selected from
the X-ray REFLEX (B\"ohringer et al.\ 2004) and eBCS catalogues (Ebeling et
al.\ 1998 and 2000), imposing the following constraints: $L_X$
(0.1--2.4~keV) $>5\times10^{44}$ erg~s$^{-1}$; $0.2 < z < 0.4$;
$-30^{\circ}<\delta<+2.5^{\circ}$ (REFLEX) and $+15^{\circ}<\delta<+60^{\circ}$
(eBCS). A number of the selected clusters had either archival or literature
information, with 7 known radio halos. With the GMRT we observed 34/50
clusters, i.e.\ those lacking high sensitivity imaging.

The whole observational project is described in Venturi et al.\ 2007 (V07) and
2008 (V08). The observations were carried out at 610~MHz; each cluster was
observed for $\sim 2.5{-}3.5$ hours, for an average noise in the images of the
order of $35{-}100$ $\muup$Jy beam$^{-1}$, mainly depending on the presence of
strong sources in the field.

A number of new diffuse cluster sources (halos, relics and mini-halos) were
found, as well as candidates (see next section and Table~1), but the great
majority of the observed clusters (25/34) do not host diffuse emission at their
centres. This result in itself confirms that radio halos are rare. Inspection
of the X-ray properties of the clusters in the sample allowed us to conclude
that: (a) all clusters hosting halos and relics show evidence of strong merger
events; (b) clusters with mini-halos have a cooling core; (c) clusters without
diffuse emission may be either dynamically active or relaxed. These results are
consistent with the expectations of the re-acceleration model.

The results of the GMRT Radio Halo Survey (both detections and upper limits)
were combined with the information from the Northern VLA Sky Survey (NVSS) for
the clusters in the redshift interval $z=0{-}0.2$ (Giovannini et al.\ 1999), so
as to have a picture of the radio halo phenomenon in the large redshift
interval $z=0{-}0.4$. It was found that galaxy clusters are either ``radio
loud'', hosting a giant halo whose power correlates with the ICM X-ray
luminosity (CBS06), or ``radio quiet'', i.e.\ they do not host diffuse central
emission down to a level which is at least one order of magnitude below the
$\log L_{\rm X}{-}\log P_{\rm 1.4~GHz}$ correlation (Brunetti et al.\ 2007).
Moreover, it was quantitatively shown that the fraction of clusters with a
giant radio halo increases with increasing cluster mass (X-ray luminosity), at
a statistically significant level (Cassano et al.\ 2008). The X-ray luminosity
threshold is $L_{\rm X} \sim 10^{45}$ erg~s$^{-1}$. All the above results are
pieces of evidence in favour of the re-acceleration model by turbulence.

\section{Low frequency follow-up of diffuse sources in the GMRT cluster sample}

Most of the radio halos and relics known to date have been detected and imaged
only at 1.4 GHz (i.e.\ Bacchi et al.\ 2003; Govoni et al.\ 2001 and 2004) and
very little information is available on their emission properties at
frequencies $\nu \le 325$~MHz. Therefore we started a GMRT low frequency
follow-up of the new cluster sources detected at 610~MHz and of the candidates
(V07 and V08), and of the radio halos in the sample with literature
information. The list of clusters observed and the frequency chosen are
reported in Table 1.

Each cluster was observed for 8 hours, recording both upper and lower side
bands. The quality of the data is generaly good; the 1$\sigma$ rms level
reached in the images is in the range $0.1{-}0.5$ mJy beam$^{-1}$, mainly
depending on the amount of editing required as consequence of RFI.

\begin{figure}
\plottwo{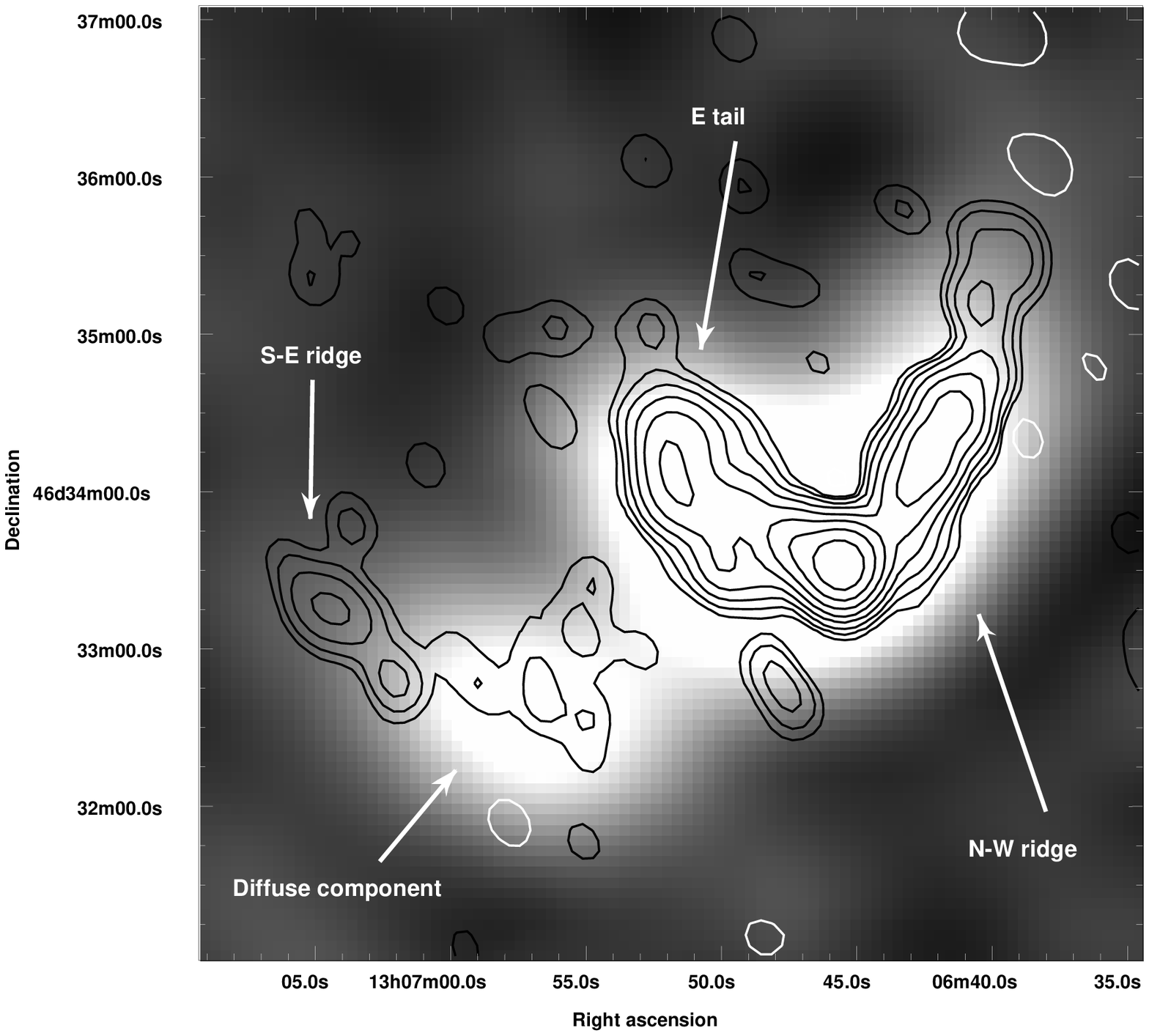}{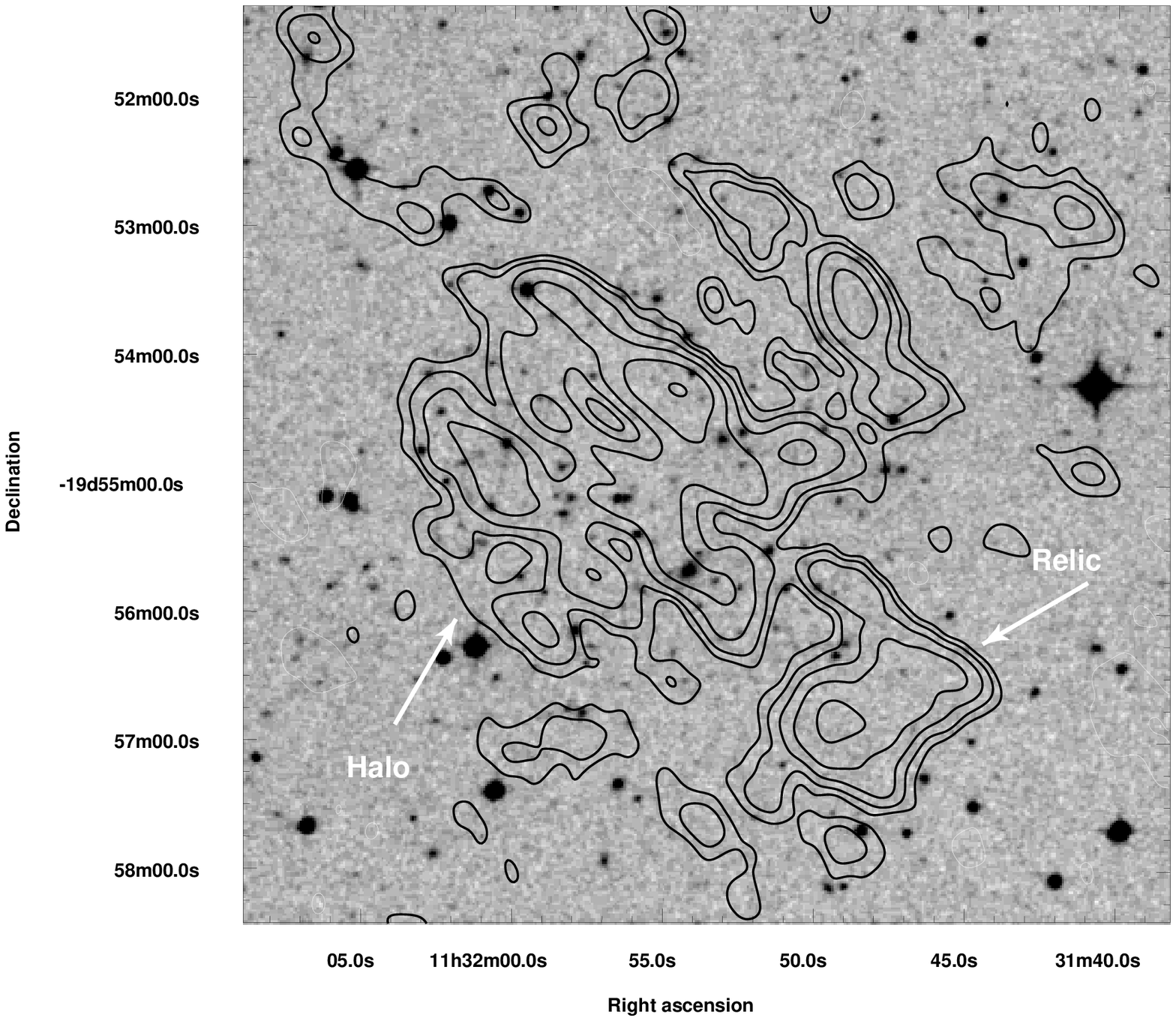}
\caption{{\itshape Left:\/} A\,1682. GMRT contours at 240~MHz overlaid on the
VLSS. Contours are $2.0 \times (-1, 1, 2, 4, 8,\dots)$ mJy beam$^{-1}$;
$1\sigma=0.6$ mJy beam$^{-1}$. The restoring beam is $18'' \times 14''$.
{\itshape Right:\/} A\,1300. GMRT contours at 325~MHz overlaid on the DSS-2.
Individual sources have been subtracted. Contours are $0.75 \times (-1, 1, 2,
4, 8,\dots)$ mJy beam$^{-1}$; $1\sigma=0.25$ mJy beam$^{-1}$. The restoring
beam is $27'' \times 16''$.}
\end{figure}

\subsection{A\,521: the first ultrasteep spectrum radio halo}

The low frequency observations of A\,521, carried out to study the cluster
relic (Giacintucci et al.\ 2008), led to the discovery of  a very steep
spectrum ($\alpha \sim 2$) giant radio halo  at the cluster centre (Brunetti et
al.\ 2008). The existence of radio halos with very steep spectrum is consistent
with the re-acceleration scenario in case of less energetic events, which would
not be able to provide enough turbulence to re-accelerate electrons at the
energies requested to have GHz emitting radio halos (see also the paper by
Cassano, this volume). Given that minor mergers are very common in the
Universe, we expect that many more ultra steep spectrum radio halos (USSRH) may
be found, when searched for with appropriate cluster selection criteria and
observing strategies.

\subsection{A\,697: a candidate USSRH}

The faint radio halo visible in this cluster at 610~MHz (V08) has a flux
density $S_{\rm 325~MHz} \sim$ 45 mJy, leading to a spectral index $\alpha_{\rm
325~MHz}^{\rm 610~MHz} \sim$ 2. The cluster largest linear size is of the order
of the Mpc, and its morphology is in very good agreement with the underlying
X-ray emission. A detailed study of this cluster is in progress (Macario et al.
to be submitted).

\subsection{A\,781: an intriguing diffuse peripheral source}

The diffuse source detected at 610~MHz (V08) is much more extended at 325~MHz,
reaching a maximum extent in the North--South direction of $\sim 1$ Mpc. The
total flux density for this source is 80 mJy. The spectral index is
$\alpha_{\rm 325~MHz}^{\rm 610~MHz} \sim 1.2$, if the flux density is
integrated over the same area. Such value is not in very good agreement with
the value  $\alpha^{1.4 GHz}_{610~MHz}$=0.78 reported in V08, and this could be
due to different areas used for the total flux density integration. Consistency
checks are in progress.

\subsection{A\,1682: a very complex cluster}

Our 240~MHz observations confirm that the radio emission in this cluster is
very complex. The left panel of Fig.~1 reports the 240~MHz radio contours
overlaid on the VLSS. We confirm that the S--E ridge has a steep spectrum:
$\alpha^{\rm 610~MHz}_{\rm 240~MHz} \sim 1.4$ ($S_{\rm 240~MHz}=59$ mJy). The
analysis of the E tail and of the N--W ridge is in progress. The most
remarkable feature in the 240~MHz image of A\,1682 is the detection of diffuse
emission with a strong counterpart on the VLSS (see Fig.~1). A detailed study
is in progress (Cassano et al.\ in prep.).

\subsection{A\,1300 and A\,2744}

Both clusters, known from the literature, host a giant radio halo and a relic.
The radio halo in A\,1300, shown in the right panel of Fig.~1, has a largest
linear extent of $\sim$ 1.3 Mpc, much larger than inferred in the MOST image
published in Reid et al.\ (1998). The flux density detected in the radio halo
and in the relic are respectively $S_{\rm 325~MHz}=290$ mJy and $S_{\rm
325~MHz}=77$ mJy. Using the flux density measurements reported in Reid et al.\
(1998) for the candidate relic we obtained $\alpha_{325~MHz}^{1.4 GHz} \sim 1$.
No estimate of the spectral index be given for the radio halo.

Thanks to the better sensitivity of our image compared to those in Orr\'u et
al.\ (2007), the radio halo in A\,2744 is larger than previously imaged at
325~MHz, with a largest linear size of the order of 1.5 Mpc. Its total flux
density is $S_{\rm 325~MHz}$=330 mJy, while for the relic we measured $S_{\rm
325~MHz}$=122 mJy.

\subsection{RXCJ\,2003.5$-$2323}

This giant radio halo (largest linear size 1.4 Mpc) was discovered with the
GMRT Radio Halo Survey and has been extensively studied in the radio, X-ray and
optical band (Giacintucci et al.\ to be submitted). It is very powerful, with a
single power law spectrum with $\alpha_{\rm 240~MHz}^{\rm 1.4 GHz} \sim 1.3$.
Our multiwavelength study supports the scenario of a merger-driven formation
for this giant radio halo.

\acknowledgements
We thank N. Kantharia and the GMRT staff for their help during the
observations. GMRT is run by the National Centre for Radio Astrophysics of the
Tata Institute of Fundamental Research. We acknowledge contribution from grants
ASI-INAF I/088/06/0 and PRIN-INAF 2007.


\begin{thebibliography}{}
%
\bibitem[]{1}Bacchi M., Feretti L., Giovannini G., Govoni, F., 2003, A\&A, 400,
465
%
\bibitem[]{2} B\"ohringer H., Schuecker P., Guzzo L., Collins, C.A., Voges, W.,
Cruddace, R.G., Ortiz-Gil, A., Chincarini, G., De Grandi, S., Edge, A.C., et
al., 2004, A\&A, 425, 367
%
\bibitem[]{3}Brunetti G., Setti G., Feretti L., Giovannini, G., 2001, MNRAS,
320, 365
%
\bibitem[]{4}Brunetti G., Venturi T., Dallacasa D., Cassano, R., Dolag, K.,
Giacintucci, S., Setti, G., 2007, ApJ, 670, L5
%
\bibitem[]{5}Brunetti G., Giacintucci S., Cassano R., Lane, W., Dallacasa, D.,
Venturi, T., Kassim, N.E., Setti, G., Cotton, W.D., Markevitch, M., 2008,
Nature, 455, 944
%
\bibitem[]{6}Cassano R., Brunetti G., 2005, MNRAS, 357, 1313
%
\bibitem[]{7}Cassano R., Brunetti G., Setti G., 2006, MNRAS, 369, 1577 (CBS06)
%
\bibitem[]{8}Cassano R., Brunetti G., Venturi T., Setti, G., Dallacasa, D.,
Giacintucci, S., Bardelli, S., 2008, A\&A, 480, 687
%
\bibitem[]{9}Ebelin H., Edge A.C., B\"ohringer H., Allen, S.W., Crawford, C.S.,
Fabian, A.C., Voges, W., Huchra, J.P., 1998, MNRAS, 301, 881
%
\bibitem[]{10}Ebelin H., Edge A.C., Allen S.W., Crawford, C.S., Fabian, A.C.,
Huchra, J.P., 2000, MNRAS, 318, 333
%
\bibitem[]{11}Feretti L., 2005, in X-ray and Radio Connections, Sjouwerman
L.O., Dyer K.K., eds, published electronically by NRAO,
{\tt http://www.aoc.nrao.edu/events/xraydio/}
%
\bibitem[]{12}Giacintucci S., Venturi ., Macario G., Dallacasa, D., Brunetti,
G., Markevitch, M., Cassano, R., Bardelli, S., Athreya, R., 2008, A\&A, 486,
347
%
\bibitem[]{13}Giovannini G., Tordi M., Feretti L., 1999, New Astron., 4, 141
%
\bibitem[]{14}Govoni F., Feretti L., Giovannini G., 2001, A\&A, 376, 803
%
\bibitem[]{15}Govoni F., Markevitch M., Vikhlinin A., Van Speybroek, L.,
Feretti, L., Giovannini, G., 2004, ApJ, 605, 695
%
\bibitem[]{16}Orr\'u E., Murgia M., Feretti L., Govoni, F., Brunetti, G.,
Giovannini, G., Girardi, M., Setti, G., 2007, A\&A, 467, 943
%
\bibitem[]{17}Petrosian V., 2001, ApJ, 577, 560
%
\bibitem[]{18}Reid A.D., Hunstead R.W., Lemonon L., Pierre M.M., 1998, MNRAS,
302,571
%
\bibitem[]{19}Venturi T., Giacintucci S., Brunetti G., Cassano, R., Bardelli,
S., Dallacasa, D., Setti, G., 2007, A\&A, 463, 937 (V07)
%
\bibitem[]{20}Venturi T., Giacintucci S., Dallacasa D., Cassano, R., Brunetti,
G., Bardelli, S., Setti, G., 2008, A\&A, 484, 327 (V08)
%
\end{thebibliography}
\end{document}